# Manipulating azobenzene photoisomerization through strong light–molecule coupling

J. Fregoni[1,2], G. Granucci[3], E. Coccia[4], M. Persico[3] & S. Corni[2,4]

The formation of hybrid light–molecule states (polaritons) offers a new strategy to manipulate the photochemistry of molecules. To fully exploit its potential, one needs to build a toolbox of polaritonic phenomenologies that supplement those of standard photochemistry. By means of a state-of-the-art computational photochemistry approach extended to the strong-coupling regime, here we disclose various mechanisms peculiar of polaritonic chemistry: coherent population oscillations between polaritons, quenching by trapping in dead-end polaritonic states and the alteration of the photochemical reaction pathway and quantum yields. We focus on azobenzene photoisomerization, that encompasses the essential features of complex photochemical reactions such as the presence of conical intersections and reaction coordinates involving multiple internal modes. In the strong coupling regime, a polaritonic conical intersection arises and we characterize its role in the photochemical process. Our chemically detailed simulations provide a framework to rationalize how the strong coupling impacts the photochemistry of realistic molecules.

[1] Dipartimento di Scienze Fisiche, Informatiche e Matematiche, University of Modena and Reggio Emilia, I-41125 Modena, Italy. [2] Istituto Nanoscienze, Consiglio Nazionale delle Ricerche CNR-NANO, I-41125 Modena, Italy. [3] Dipartimento di Chimica e Chimica Industriale, University of Pisa, I-56124 Pisa, Italy. [4] Dipartimento di Scienze Chimiche, University of Padova, I-35131 Padova, Italy. Correspondence and requests for materials should be addressed to G.G. (email: giovanni.granucci@unipi.it) or to S.C. (email: stefano.corni@unipd.it)

Control and manipulation of the photochemistry of molecules has traditionally relied on synthetic[1] chemical modifications or changes in the environment surrounding the photoactive molecule[2]. Precise control of the main products, the reaction yields and rates is achievable through addition or removal of functional groups that modify the properties of the ground and excited states. Together with the idea that light–matter interaction in both the weak-field and strong-field limits can be exploited to control molecular processes[3–7], it has been suggested recently that the light–molecule interaction itself can be used to modify the photochemistry of the molecules, with no other direct changes in the molecule or its environment[8–10]. We specifically refer to the regime where the coherent energy exchange rate $g$ (also addressed as a coupling constant) between light and molecules becomes faster than any decay rate of the system itself (strong-coupling limit) (Fig. 1a). In this regime, the states of the system become hybrid between light and matter, the so-called polaritons[11–13].

Such states mix the photonic and the electronic degrees of freedom: when compared to the pure electronic states, the properties of the polaritons show a different dependence on the molecular geometry. This applies, in particular, to the polaritonic potential energy surfaces (PPESs) and any feature related to them, such as avoided crossings and conical intersections. As a consequence, the polaritons may impart a new photochemistry, laying the basis for polaritonic chemistry[8,9,13–16].

To achieve the strong coupling required to exploit such hybrid states, resonant or nanoplasmonic cavities have been devised only in the last years[17,18] and recently the single-molecule level has been reached[11,19] at room temperature. Coherent coupling between a single organic molecule and a microcavity has also been recently achieved[20], opening a way to the investigation of coherence effects in the light–matter interactions on a longer timescale than in the nanocavity case. Such groundbreaking experimental findings have spurred an intense and pioneering theoretical activity[12,14,21–23]. Various interesting phenomena were predicted in prototypical systems (model potential energy surfaces—PESs—along one or two coordinates, representing specific internal coordinates in more complex systems), such as the modification of quantum yields and the creation of polaritonic conical intersections by light–molecule coupling[13,15,21,24,25].

The computational investigation of weak-field photochemistry[5,6,26–29] has undoubtedly shown that even for the simplest and best characterized systems, such as azobenzene, the chemical complexity of the molecule cannot be disregarded. To tackle such complexity, the inclusion of all the molecular degrees of freedom is necessary to describe the main features of photochemical processes: the occurring events where the Born–Oppenheimer approximation breaks down and the correct account of the quantum nature of the nuclear motion. In the past few decades, many efforts in this field resulted in detailed and realistic models[30,31] of photoactive systems.

An equivalent investigation of the photochemical properties and peculiar features of realistic molecules in the strong-coupling regime is still an open challenge[14]. Recently, methodologically remarkable advancements have been made along such direction: Luk et al. investigated the formation of collective polaritonic states for systems of hundreds of realistic dye molecules[22], and Vendrell focused on collective quantum effects for up to five diatomic molecules strongly coupled to a cavity mode[23]. Yet, the characterization of a photochemical reaction in the strong-coupling regime for a molecule of realistic complexity is still lacking. To move further in the characterization of photochemical processes, we focus on azobenzene. Azobenzene and its derivatives are a prototypical benchmark for studying photochemical processes[32,33]. The photo-reversible switch of configuration between *trans* and *cis* (Fig. 1b) in this class of molecules has been studied extensively, due to the wide applicability in the field of photocontrol of biomolecular structures[34], of sensing[35], and photoresponsive materials[36].

To investigate such system, we rely on an on-the-fly surface hopping technique[37] already validated for several applications[31,38,39]. Here we characterize the PPESs by relying on the detailed description of the molecule taking into account the full space of the internal degrees of freedom of azobenzene. We make use of such characterization to discuss the effects of strong coupling on photochemistry, as the birth of coherent oscillations of the populations between the polaritonic states. For coupling strengths comparable to what was already achieved in experiments[11], we show how the mechanism of the *trans*–*cis* photoisomerization is modified, leading to a decrease in the quantum yield[24,25]. Finally, we investigate the oscillations referred above by including the effect of cavity losses to mimic realistic plasmonic nanocavities. Through our results, we emphasize the significant role of quantum coherence in controlling the molecular processes, including in the picture also the polaritonic coherences beside electronic and vibrational ones[29,40,41].

## Results

**Azobenzene polaritonic PESs**. The azobenzene electronic PESs are represented with respect to the *CNNC* and one of the *NNC*

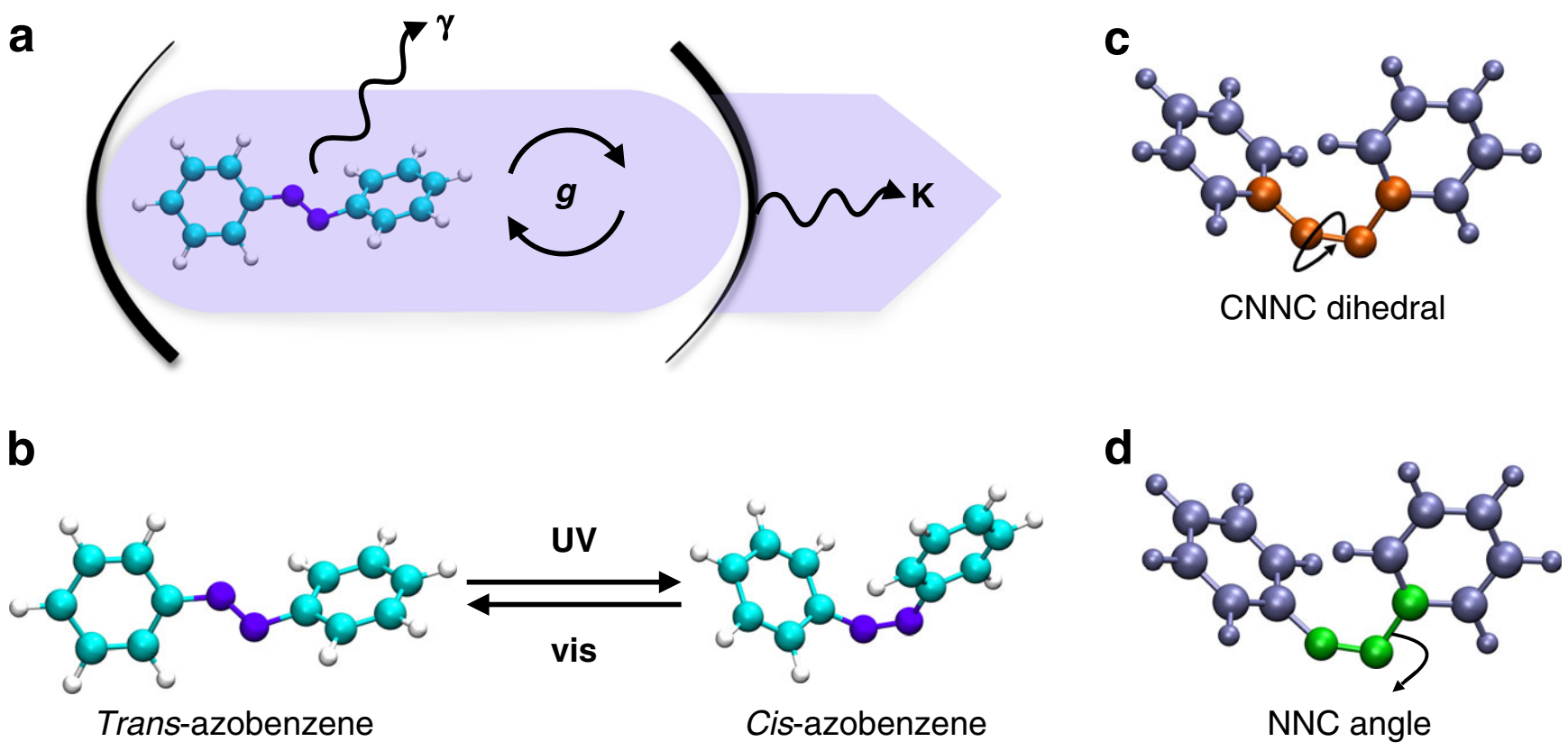

**Fig. 1** Scheme of the modeled system. **a** *Trans* azobenzene molecule inside a resonant cavity. The decay rate of the system is $\gamma$ for the molecule and $\kappa$ for the cavity. **b** Isomerization of the azobenzene molecule: the switch of configuration can be achieved by irradiation with UV–Vis light. **c**, **d** Photoisomerization coordinates of azobenzene: the reaction occurs through the torsion of the *CNNC* dihedral angle, with a simultaneous relaxation of the *NNC* angle

angular coordinates (Fig. 1c, d) with all the other nuclear degrees of freedom optimized with respect to the first excited state and for each choice of such coordinates. These two angles are directly involved in the photoisomerization mechanism[42]. The *CNNC* dihedral angle (Fig. 1c) represents the torsion around the N=N double bond and it is the main reaction coordinate for the photoisomerization. Together with the torsion, the *NNC* angles are necessary to successfully describe the photoisomerization mechanism[38,42,43]. Electronic wavefunctions and PESs (Fig. 2a) were computed exploiting a semiempirical quantum chemistry approach, developed by Persico and collaborators[38,43]. Such approach has been extensively validated in the past for azobenzene and its derivatives[38,43,44]. Accuracy and low computational cost make this method extremely suitable to simulate the photochemistry of realistic molecules, as it allows the inclusion of all the internal degrees of freedom.

We build (see Methods) the polaritonic states as eigenstates of the total (molecule + light) Hamiltonian on the basis of product states between the electronic eigenstates $S_0$, $S_1$, and a photon occupation state number $|0\rangle$, $|1\rangle$, that is $|S_0, 0\rangle$, $|S_1, 0\rangle$, $|S_0, 1\rangle$, and $|S_1, 1\rangle$. The strong-coupling interaction only mixes states differing by one in the photon occupation state number. The mixing between $|S_0, 1\rangle$ and $|S_1, 0\rangle$ is by far the most relevant, as these two states are close in energy. In particular, such mixing gives rise to the lower and upper polaritons ($|-\rangle$ and $|+\rangle$, respectively, see Fig. 2b)[12,21,24,45].

In Fig. 2b, they have been obtained with a coupling constant (see Methods) $g$ of 0.010 au and a photon energy $E_{\text{ph}}$ of 1.3 eV. As shown there, a new avoided crossing arises for the polaritonic states as a signature of the coupling. Such avoided crossing is found in the coordinate range where the $|S_1, 0\rangle$ and $|S_0, 1\rangle$ states would cross. The energy splitting contribution along the avoided crossing line between the $|-\rangle$ and $|+\rangle$ states is proportional to the coupling between $|S_0, 1\rangle$ and $|S_1, 0\rangle$ before the diagonalization. Such coupling depends on the expectation value of the component of the transition dipole moment between the pure electronic states, $\boldsymbol{\mu}_{S_0,S_1}(Q)$, taken along the polarization direction of the electric field (see Methods). Therefore, the splitting magnitude depends indirectly on the nuclear coordinates through the transition dipole moment. In turn, since the geometry where the strong-coupling avoided crossing occurs is tuned by $E_{\text{ph}}$, different splitting energies are obtained as a function of $E_{\text{ph}}$. Such dependence has been noted in previous works and included in the models[21,22,24,25], though its role for the photochemistry of realistic molecules has not been explored yet. By showing in Supplementary Figs. 1 and 2 that the effect of such dependence is remarkable, we anticipate that the transition dipole moment at a given crossing geometry (governed by $E_{ph}$) is a further parameter to take into account to manipulate the PPESs features and, as a consequence, the photochemical reaction. The splitting can range from zero (for dipole forbidden transitions) to very large, depending on the transition dipole magnitude and orientation.

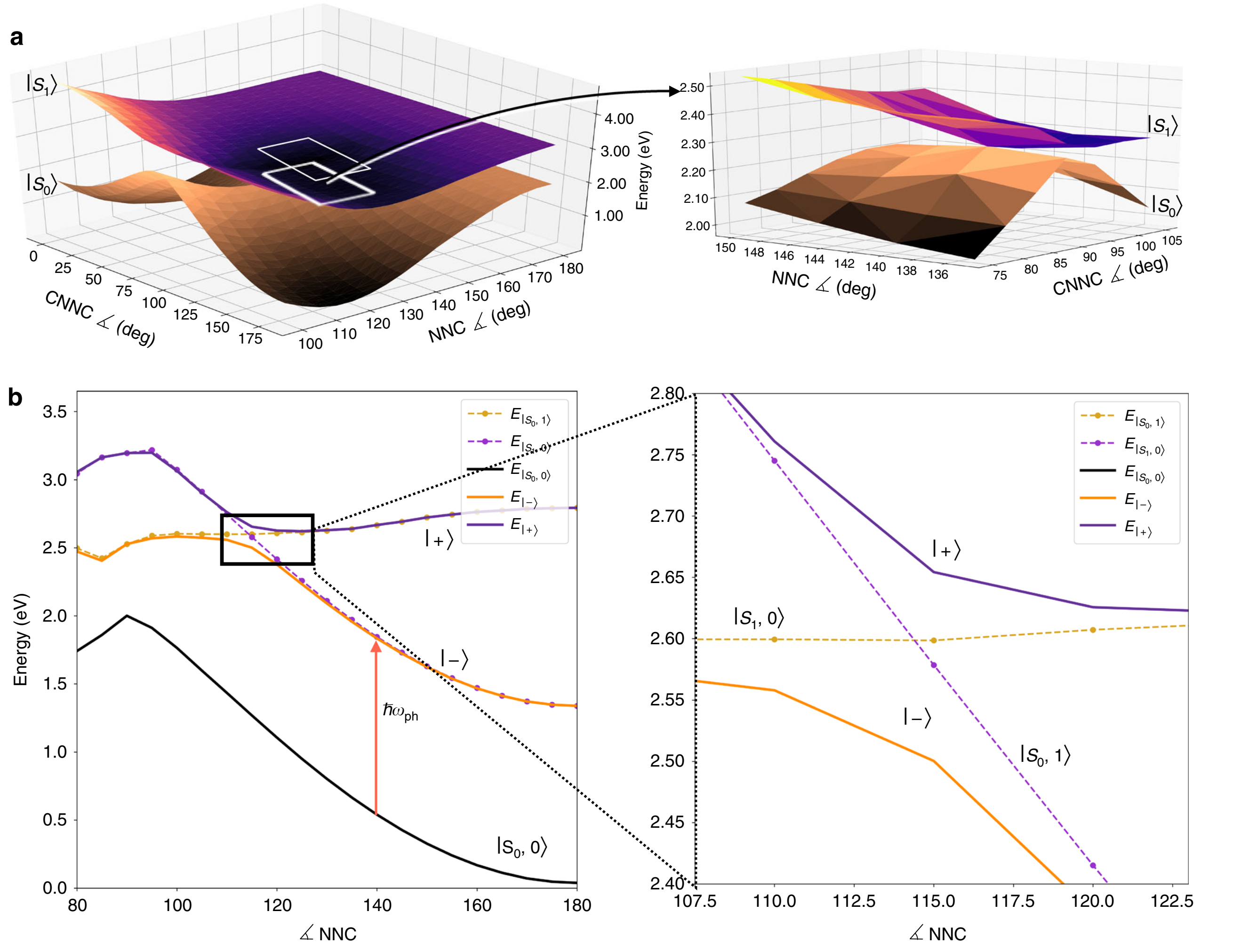

**Fig. 2** Electronic and polaritonic states of azobenzene. **a** Pure electronic ground (brown surface) and first excited (purple surface) PESs of azobenzene, plotted with respect to the torsion and inversion coordinates. The $S_0$ and $S_1$ PESs are characterized by the presence of a conical intersection. **b** Polaritonic potential energy curves (black, dark orange, and purple full lines) of azobenzene with respect to the *CNNC* coordinate with *NNC* 115°, obtained as linear combinations of uncoupled states (dotted lines, orange and violet) for a photon energy $E_{\text{ph}}$ of 1.3 eV and a coupling constant $g$ of 0.010 au. The splitting between the polaritonic states ($|-\rangle$, $|+\rangle$) depends on the transition dipole moment between the electronic states, evaluated at the crossing geometry (*CNNC* 130° and *NNC* 115°)

The out-of-plane component of the transition dipole moment between $S_0$, and $S_1$ vanishes at planar geometries. Therefore, for a field polarization perpendicular to the plane of the molecule as in the present case, the $|-\rangle$ and $|+\rangle$ states become exactly degenerate at crossing points between $|S_1, 0\rangle$ and $|S_0, 1\rangle$ at planar geometries. In other words, a polaritonic conical intersection (indicated with a red arrow in Figs. 3, 4a and 4d) is originated (see Supplementary Note 1 for more details).

Aiming to highlight how deeply the conical intersection features can influence the photoisomerization yields and mechanism, we chose two limiting cases which shape very differently the PPESs. Within such two cases, the photon energy $E_{ph}$ is set to 1.3 and 2.2 eV, while $g$ is equal to 0.010 au and the field is polarized perpendicularly to the plane of the molecule for both. By doing so, we obtain coupling magnitudes which are comparable to what has recently been observed experimentally for single molecules in the strong-coupling regime[11,19]. In the next section, we analyze the differences between strong coupling-induced avoided crossings under different conditions and discuss the consequences of the photoisomerization process of azobenzene.

**Photochemistry in the strong-coupling regime**. The simulation of a photochemical process is carried out by non-adiabatic molecular dynamics methods: this manifold of techniques[37,46–48] consists in mimicking the nuclear wavepacket motion on the excited electronic PESs and aims to correctly retrieve the quantum yields of a reaction when the Born–Oppenheimer approximation breaks down. Such breaking can occur either for electronic states degeneracy, quasi degeneracy[49,50] or, in the strong-coupling regime we are considering, for polaritonic state avoided crossings[12,25,51]. In these critical regions, reproducing the correct splitting of the wavepacket through crossing seams is the key to correctly retrieve the molecular mechanism[30,44,49,50,52].

To this purpose, an effective[46,47] strategy is to rely on the semiclassical surface hopping method pioneered by Tully[53]. Such framework provides an accurate description of de-excitation mechanisms in molecules. A recent improvement to such an approach includes decoherence effects[54], which have been proven successful to describe multiple passages of the wavepacket through the crossing seams. The accurate description of multiple passages is essential to our system: the presence of the conical intersection and the strong-coupling avoided crossing entails multiple wavepacket branchings in rapid succession. Therefore, we have devised a propagation scheme for the nuclear trajectories on the PPESs, in the framework of an on-the-fly trajectory surface hopping technique[37] (see Methods for more details).

As mentioned above, we compare two cases with different photon energies ($E_{ph} = 1.3$ eV and 2.2 eV), $g$ equal to 0.010 au and the field polarized along the $z$-axis perpendicular to the plane of the molecule for both. The initial conditions (nuclear coordinates and momenta) for the swarm of trajectories mimicking the nuclear wavepacket are sampled from a room-temperature Boltzmann distribution, obtained from a single trajectory propagated for 10 ps on the ground state with a Bussi–Parrinello thermostat[55]. A vertical excitation is performed to the upper polariton for each trajectory (Fig. 3a). The PPESs and some snapshots of the 1.3 eV dynamics are shown in Fig. 3 (for the 2.2 eV case and the movies of the whole dynamics, see Supplementary Movies 1 and 2). In order to compare the effect of different coupling conditions on the photochemical process, the population evolution and the characterization of the polaritons are shown for the two cases in Fig. 4.

In both cases, the vertical excitation brings the sampled trajectories on a slope of the upper PPES. As shown for the 1.3 eV case at 10 fs, the trajectories start propagating, accumulating kinetic energy toward the minimum of the upper polariton (Fig. 3a). At 20 fs, as the ensemble approaches the strong-coupling avoided crossing as well as the related polaritonic conical intersection, the trajectories split and start oscillating between the $|-\rangle$ and $|+\rangle$ polaritonic states (Fig. 3b, c). At 30 fs,

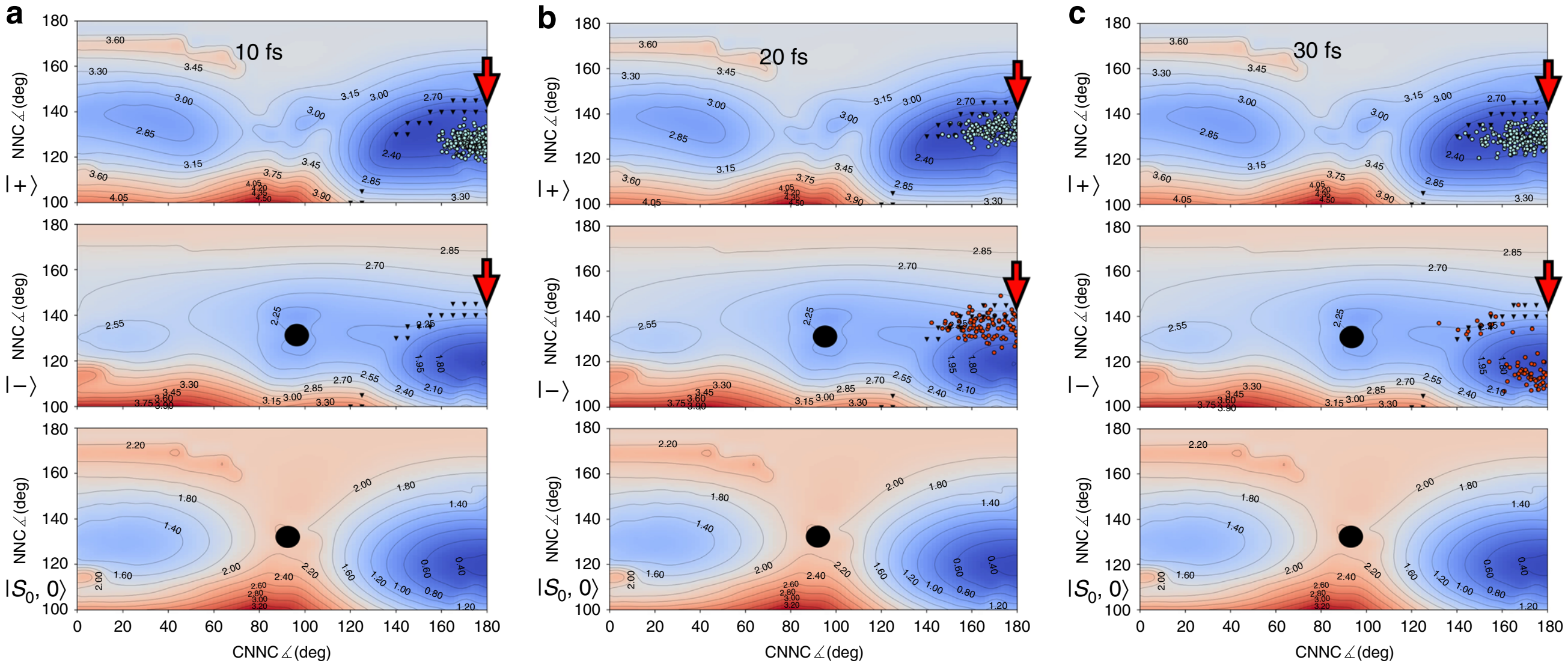

**Fig. 3** Snapshots of the photoisomerization on PPESs for $E_{ph}$ 1.3 eV and $g$ 0.010 au. The black triangles identify the strong-coupling avoided crossing line, the red arrow identifies the polaritonic conical intersection while the black circle is the conical intersection between the pure electronic states. Light blue trajectories are referred to the upper polaritons and orange trajectories are referred to the lower polariton. **a** Upon the vertical excitation, the ensemble of trajectories is found in a high slope region of the upper PPES. **b** The trajectories move toward the upper polariton minimum and **c** start oscillating between the upper and lower polaritons, with several trajectories trapped in the lower polariton minimum after the splitting (see Supplementary Movie 1)

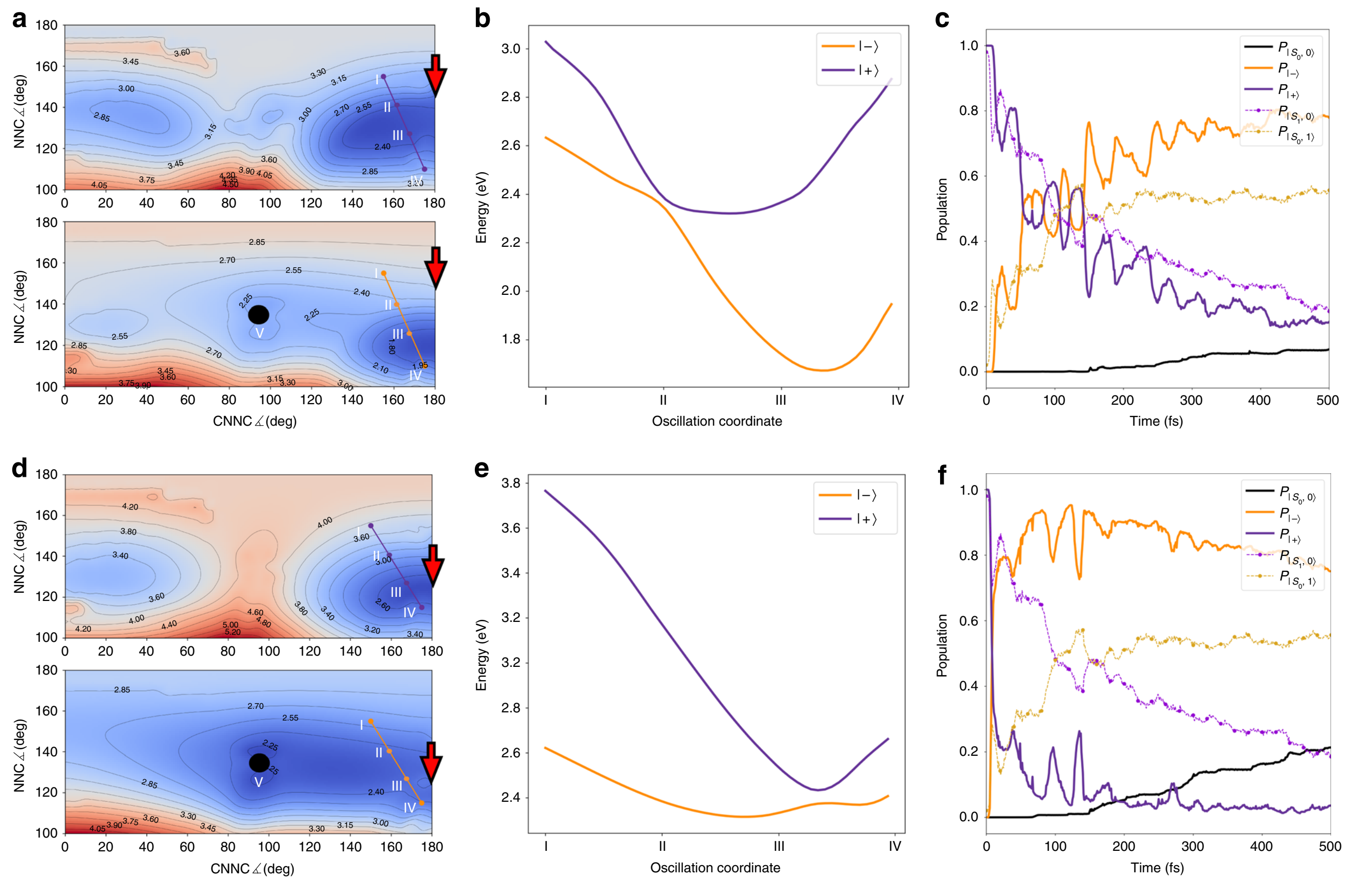

**Fig. 4** Non-adiabatic dynamics in the strong coupling regime. **a**, **b**, **d**, **e** Potential energy surfaces and state populations. The upper (respectively, lower) panels refer to photon energy $E_{\mathrm{ph}}$ 1.3 eV (respectively, 2.2 eV). In both cases, we set the coupling constant $g$ to 0.010 au and field polarized along $z$. Contour maps of the $|-\rangle$ and $|+\rangle$ PPESs are shown in **a** and **d**, while the section along the oscillation coordinate (see text) is shown in **b** and **e**. The red arrow identifies the polaritonic conical intersection, points I and IV identify the extrema of oscillations (see text), and point V indicates the electronic conical intersection. **c**, **f** Time evolution of the state populations. Full lines are referred to polaritonic states and dashed lines to the uncoupled states, according to the color scheme defined in Fig. 2

we observe a branching of the trajectories on the lower PPES, with a cluster of trajectories moving toward the lower PPES minimum (Fig. 3c).

Such oscillations were characterized for both the 1.3 and 2.2 eV cases, as shown in the population plot reported in Fig. 4. To this aim, we extracted the oscillation coordinate by averaging the *NNC* and *CNNC* angles for the oscillating trajectories, and we plotted the avoided crossing profiles for both the dynamics (Fig. 4a, b, d, e) along such a coordinate. The extrema of the (*CNNC*, *NNC*) couple of angles during the oscillations are indicated as I and IV in the figure. In both cases, we observe that the oscillation coordinate is hybrid between *CNNC* and *NNC*, though with a different mixing of the two. We also stress the different shape of the surfaces and of the crossing profiles between the polaritonic states in the two cases, despite the same coupling constant. The splitting extent at the avoided crossing (hence, proportional to the coupling) is a signature of the geometric dependence of the coupling through the transition dipole moment. This dependence carries non-trivial effects on the dynamics and the oscillation feature, as it becomes clear by analyzing the populations of the two cases reported in Fig. 4c, f. The different decay rates of the upper polaritonic state and the oscillation peaks strengths and occurrences in time can be rationalized by exploring further the potential energy curves along the oscillation coordinate.

In the 1.3 eV case (Fig. 4c), the excited trajectories starting in the Franck–Condon region on the upper PPES (close to point IV in Fig. 4a) are dragged toward a strong-interaction region between the two polaritonic states. As a consequence, a diabatic behavior is obtained, with the trajectories oscillating on the $|S_1, 0\rangle$ PES between the points I and IV (Fig. 4b). However, during the oscillations, some trajectories get trapped in the minimum region of the lower polariton, which becomes therefore more and more populated (see Supplementary Fig. 3 for a sketch representation). In this specific case, a motion toward the conical intersection (point V, Fig. 4a) becomes unfavorable due to the oscillation trap triggered by the peculiar shaping of the PESs. This process occurs completely in the *trans* region of the conformational space: as a consequence, an almost complete quenching of the reaction is observed (the quantum yield is reduced to 3.1%). The 2.2 eV case is substantially different: the wavepacket starts its propagation close to the avoided crossing located at the point II of Fig. 4e, and 4f, entailing the quick upper polariton population drop reported in Fig. 4f.

In this case, the upper PES has a minimum coincident with the strong-coupling avoided crossing: while few trajectories oscillate due to the coupling effects, the wavepacket can evolve toward the conical intersection on the lower polariton (point V, Fig. 4d), damping the oscillation trap effect and resulting in a quantum yield of 16.1% (going toward the 33.1% for the isolated molecule). We also observe an effect on the populations driven by the different PPESs shapes in the two cases. For $E_{\mathrm{ph}}$ 1.3 eV, the $|-\rangle$ state has a deeper minimum, which is located far from the avoided crossing. As a consequence, the polaritonic and uncoupled populations tend to coincide quicker than in the 2.2 eV case. Due to the trajectories falling in the $|-\rangle$ PPES minimum (coincident with the $|S_0, 1\rangle$ state in that region), the $|S_0, 1\rangle$ state population grows significantly higher than the $|S_1, 0\rangle$ population. In the 2.2 eV case, the presence of a shallow minimum close to the avoided crossing entails that the trajectories are located in a region where the $|S_0, 1\rangle$ and $|S_1, 0\rangle$ states are very mixed. This behavior is well shown in Fig. 4f by the remarkable difference between the polaritonic and uncoupled states populations. It is clear from Fig.4a, d that the oscillation coordinate involves the polaritonic conical intersection neither for the 1.3 eV nor for the 2.2 eV case. Yet, inspection of the trajectory swarm shows that in both cases, some trajectories do approach the polaritonic conical intersection (and turn to the lower PPES there), with a higher probability for the 2.2 eV case due to the local shape of the PPES.

Comparing the presented results with the weak-coupling case (see Supplementary Movie 3), the pathway followed by the swarm of trajectories is substantially different. The presence of the trapping minimum in the polaritonic case (absent in the isolated molecule) limits the motion along the *CNNC* coordinate. As a consequence, the torsional photoisomerization mechanism turns out to be quenched. In addition, the oscillations along the *NNC* coordinate, inducing the periodic crossing of the polaritonic conical intersection region, not only elicit the oscillations of the polaritonic populations discussed so far: indeed, they provide a channel to intermittently fall in the electronic ground state (actually $|S_1, 0\rangle$), which is fully missing in the weak-coupling regime. The effect of such process on the ground-state population retrieval is discussed in the next section.

Finally, we evaluated the quantum yield as a function of the photon frequency in the strong-coupling regime. The trend, that is presented and discussed in the Supplementary Note 2, is non-trivial. In particular, it encompasses three different regions (a decrease, a plateau, and a recovery, reaching even a modest improvement of the yield with respect to the weak-coupling result), that testify the complexity of the strong-coupling effects on photochemistry.

**Effect of cavity losses on the photochemistry**. So far, we have presented results that explicitly account for the decay of excited electronic states. Nevertheless, they did not take into account the finite lifetime of the photon in the resonant cavity driven by the unavoidable cavity losses. While a promising coherent energy exchange has been recently shown for a single molecule within a high-quality factor (i.e., low losses) microcavity[20], so far lossy plasmonic nanocavities have been used to achieve strong coupling at the single-molecule level[11,17–19]. The electromagnetic excitation of such systems is typically characterized by a lifetime of few tens of femtoseconds, i.e., on the same timescale of the coherent oscillations described in previous sections (Fig. 4c, f). As a consequence, the lifetime of the electromagnetic excitation cannot be neglected.

In this section, we investigate the consequences of the cavity losses by means of a Monte Carlo approach (see Methods). In particular, we check whether the observed coherent oscillations persist or are dismantled by the loss of photon. We hereby consider photon lifetimes $\tau_c = 1/\kappa$ of 10 fs, 50 fs, 100 fs (Fig. 5a–c, respectively), and 150 fs to investigate their impact on the 1.3 eV dynamics (see Supplementary Note 3 for the 2.2 eV case). The main finding here is that, regardless of the specific value of the cavity lifetime used in our analysis, oscillations of the polaritonic states are retained.

To explain the persistence of such oscillations between polaritons for each investigated lifetime, it is useful to consider how the photon loss probability is determined. Akin to ref. [22], the probability of disappearance of the photon at a given time is proportional to the probability of the system to be in the state $|S_0, 1\rangle$ (the photon loss collapses the state on $|S_0, 0\rangle$). Therefore, during the time intervals in which such a probability is low, the system is protected against photon loss (i.e., the effective decay rate is much slower than $\tau_c$). On the contrary, when the state is predominantly $|S_0, 1\rangle$, such state will decay exponentially with a rate close to $\kappa$. This is evident for $\tau_c$ 10 fs (Fig. 5): every time the mimicked nuclear wavepacket passes through the polaritonic avoided crossing and $|S_1, 0\rangle$ converts to $|S_0, 1\rangle$, a clear decay with a time constant in the tens of femtoseconds range is visible. The protection against losses offered by disguising the photon in the $|S_1, 0\rangle$ state makes the dynamical features, observed above, robust in the range of hundreds of femtoseconds, despite a photon lifetime of 10 fs only.

Remarkably, the excited states oscillatory behavior translates into an oscillating probability of the molecule of being in the electronic ground state, as shown in Fig. 5d. Oscillations are clearly visible for $\tau_c$ 50 fs, and even for a limit value of $\tau_c$ 10 fs (rather short even for plasmonic nanocavities) clear periodic plateaus are visible. Current ultrafast optical experiments provide viable time resolution to observe such features, found missing in the weak-coupling regime (No SC) in Fig. 5d. Incidentally, such a panel also illustrates well the different mechanism of the reaction in the strong-coupling regime vs. weak coupling: in the former case, the ground state starts to be populated at the very beginning of the simulation due to the change in polaritonic nature upon traversing the polaritonic avoided crossing (one can interpret this as an enhanced radiative decay); in the latter, no decay takes place until the molecule reaches the electronic conical intersection (that requires around 150 fs).

Although the qualitative features are conserved, the introduction of the cavity losses does affect the results: the trapping into the lowest polaritonic state is now only transient, as it evolves toward the $|S_0, 0\rangle$ state. Moreover, the decay rate to the ground state is overall (moderately) increasing by decreasing $\tau_c$: this behavior is expected as, once $|S_0, 0\rangle$ is reached, going back to $|S_1, 0\rangle$ would require a thermal activated event. Yet, the features of the strong-coupling regime are clearly visible for all the $\tau_c$. The case of $E_{ph}$ 2.2 eV is discussed in the Supplementary Note 3; the notable point there is that for the lowest $\tau_c$, only the polaritonic state that coincides with $|S_1, 0\rangle$ survives. As a consequence, the difference between the population of polaritons and the population of the uncoupled states, shown in Fig. 4f, is also suppressed, indicating the loss of coherence between the electronic and photonic states in this very lossy cavity regime.

## Discussion

Building upon well-established methods to simulate photochemistry, we have characterized the PPESs for a realistic molecule. The simulation of the photoisomerization has shown how the reaction pathway and quantum yields can be modified in the strong-coupling regime (see Supplementary Movies 1 and 2). Tunable parameters are the coupling constant, the field mode polarization, and the resonant photon frequency, carrying strong consequences on the polaritonic splitting. In particular, the resonant frequency affects the photochemistry both by the positioning of the polaritonic avoided crossing and via the dependence of the local transition dipole moment on the avoided crossing geometry, offering a powerful (but difficult to set) handle to affect photochemistry. We have highlighted and characterized the peculiar population oscillations arising in strongly coupled azobenzene photoisomerization, promising to be probed by experimental ultrafast spectroscopy (whose role in probing polaritonic photochemistry has been already underlined)[21]. We have shown that such features are qualitatively conserved even when fast photon losses in the cavity are accounted for (much faster than the time span of the dynamics), and explained why this is the case. In addition, we have calculated and commented the non-trivial dependence of the quantum yields on different photon energies and coupling constants.

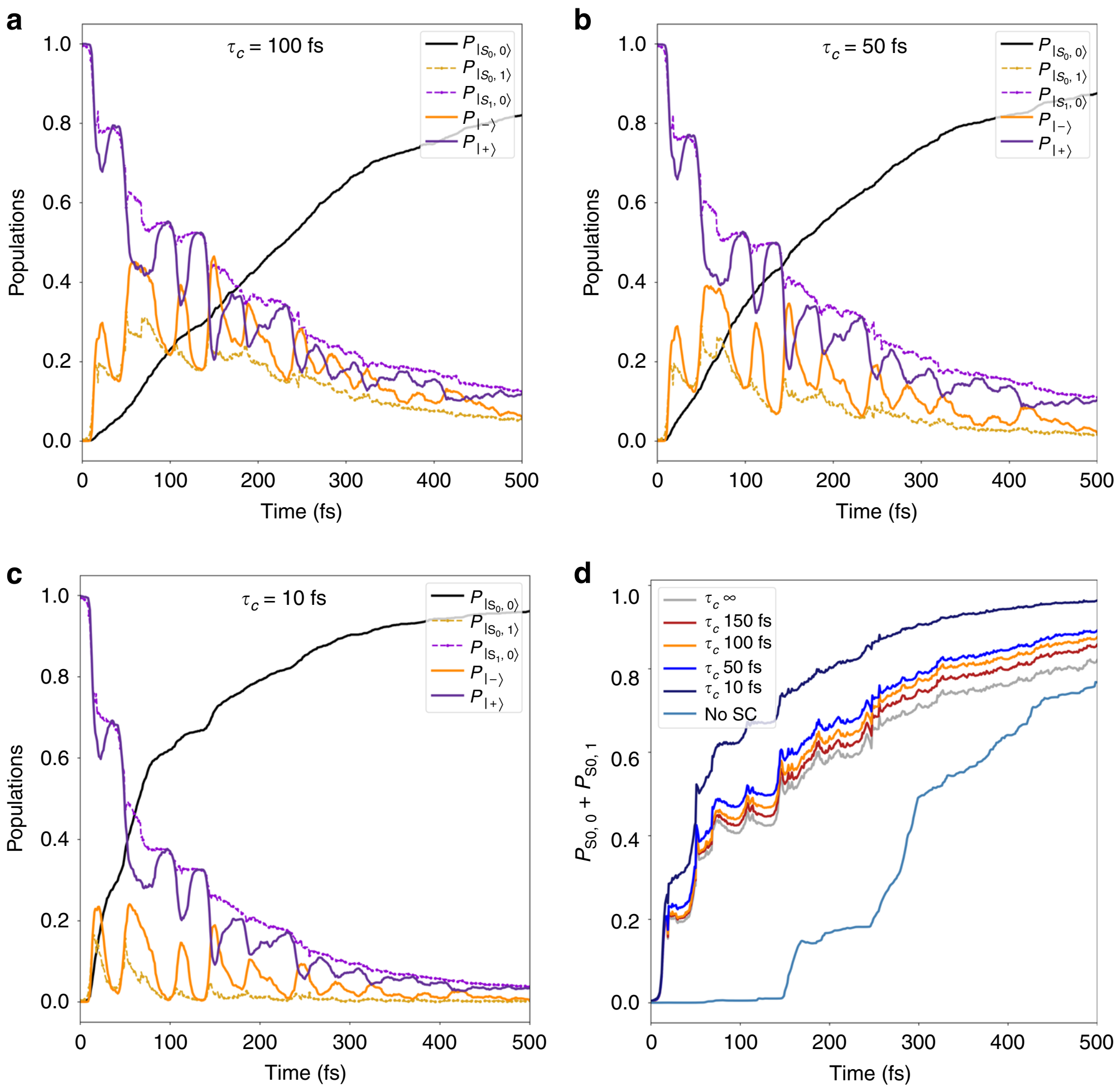

**Fig. 5** Effect of the cavity losses on the dynamics. **a–c** Full lines are referred to polaritonic states and dashed lines to the uncoupled states, according to the color scheme defined for Fig. 2. Evolution of polaritonic populations for the 1.3 eV dynamics (see Fig. 4a–c) including cavity lifetimes of 100, 50, and 10 fs, respectively. **d** Retrieval of ground-state population for different cavity lifetimes compared to the purely electronic (No SC, light blue) and the cavity lossless cases (gray)

Our results open the way to the rational design of polariton-induced control of the molecular photochemistry as, for instance, high-throughput computational investigations of the optical parameter space (photon energy, coupling strength, and electric field polarization) can be performed to find out how to control photochemistry. Among the unexplored photochemical features, the strong-coupling conditions can potentially be tuned to retrieve quantum yield enhancements instead of quenching, to maximize the photostationary state concentrations of reactant and product, to maximize the coherent oscillation aspects of the reaction mechanisms, and to tune PESs features such as the position of the electron and polaritonic conical intersections, possibly to engineer new photochemical reaction pathways.

## Methods

**Electronic states** . The electronic calculations are performed in a semiempirical framework, with an AM1 Hamiltonian, which was carefully reparameterized for azobenzene in a previous work[43]. In particular, the FOMO-CI[37,43] method has been used for the evaluation of electronic energies, wavefunctions, and couplings.

**Polaritonic states**. We rely on recent theoretical developments for strongly coupled systems[24,25,51] to build the polaritonic states. The total Hamiltonian for the system is composed by three different contributions[12,21,24]: molecule, electromagnetic field, and coupling,

$$\hat{H}_{\rm sc} = \hat{H}_{\rm mol} + \hat{H}_{\rm ph} + \hat{H}_{\rm int}. \tag{1}$$

A single quantized mode for the electromagnetic field is considered[56]:

$$\hat{H}_{\rm ph} = \omega_{\rm ph}\left(\hat{b}^{\dagger}\hat{b} + \frac{1}{2}\right), \tag{2}$$

where $\omega_{\rm ph}$ is the resonant photon frequency and $\hat{b}^{\dagger}$, $\hat{b}$ are the creation and annihilation operators for the bosonic mode, respectively. This term represents the light mode confined in resonant cavities or nanocavities. The uncoupled system, whose Hamiltonian reads $\hat{H}_{\rm mol} + \hat{H}_{\rm ph}$, is described by the product states $|S_0, 0\rangle$, $|S_1, 0\rangle$ (molecular electronic states with no photon) and the same electronic states with a photon present, $|S_0, 1\rangle$, $|S_1, 1\rangle$. The positioning of the crossing between the uncoupled states $|S_0, 1\rangle$, $|S_1, 0\rangle$ is governed by the confined mode frequency only. The PPESs are obtained by including light–matter interaction in the Coulomb gauge with a dipolar light–matter Hamiltonian (the dipolar formulation allows to exploit the molecular quantities as computed by quantum chemistry calculations):

$$\hat{H}_{\rm int}(Q) = g\hat{\mu}_{S_0,S_1}(Q)\cdot\boldsymbol{\lambda}(\hat{\boldsymbol{b}}^{\dagger} + \hat{\boldsymbol{b}}), \tag{3}$$

where $\hat{\mu}_{S_0,S_1}(Q)$ is the transition dipole moment between the electronic states at given nuclear coordinates $Q$, while $\lambda$ is the field polarization vector. Within our treatment, we include the counter rotating terms usually disregarded in the Jaynes–Cummings model, which account for the Lamb shift of the $|S_0, 0\rangle$, $|S_1, 1\rangle$ states. As the total Hamiltonian is diagonalized, the polaritonic states ($|S_0, 0\rangle$, $|-\rangle$, $|+\rangle$, $|S_1, 1\rangle$ for the present case) are obtained as linear combinations of the uncoupled states. The $|-\rangle$ and $|+\rangle$ states are the minus and plus combination between the $|S_0, 1\rangle$ and $|S_1, 0\rangle$ states, respectively. It is worth noticing that the $|S_0, 0\rangle$ and $|S_1, 1\rangle$ states are coupled as well with the other states, as a result of the diagonalization. In the present case, they are labeled as the uncoupled states since their mixing with other states is negligible for each geometry.

**About dipolar formulation**. To compute the polaritonic states, we work in the Coulomb gauge with an extended Jaynes–Cummings Hamiltonian (see Supplementary Note 4). A recent discussion about the gauge choice to treat strong-coupled systems has been proposed by Flick et al.[12], where they show that for low photon frequencies and high field intensities, a complete dipolar formulation (or minimal coupling as well) is needed[45]. To this aim, we numerically tested the complete dipolar, minimal coupling and extended Jaynes–Cummings formulations for the computation of polaritonic states. Within the resonant frequencies and field strength investigated in this work, we proved numerically that the additional terms of a complete dipolar formulation can be disregarded (Supplementary Fig. 6). Therefore, we restrain the treatment to an extended Jaynes–Cummings formulation.

**Strong-coupling non-adiabatic dynamics**. In the trajectory surface hopping framework, the nuclear wavepacket motion on the electronic PESs is mimicked by a swarm of independent classical nuclear trajectories. The electronic wavefunction is propagated on-the-fly, according to the time-dependent Schrodinger equation. The propagation of the wavefunction is carried on the adiabatic basis of the polaritonic states. As a consequence, the non-adiabatic couplings between polaritonic states take into account both the couplings between electrons and nuclei and between photon and electrons. The integration of the TDSE is performed using the local diabatization (LD) scheme[37], and the transition probabilities between polaritonic states are computed according to Tully's fewest switches algorithm[53], as adapted to local diabatization. In this version, the decoherence corrections are included as presented in ref. [54].

The classical nuclear trajectories are evolved according to Newton's equation of motion. The force acting on each atom is given by the gradient of the adiabatic polaritonic state energy. The gradients for the pure electronic states are evaluated as reported in ref. [57]. In order to evaluate the gradients arising from the light–molecule interaction contribution to the energy, we rely on the scheme proposed for spin orbit coupled systems, presented in ref. [58] (see Supplementary Methods for the flowchart).

**Cavity losses**. The cavity losses are included through a Monte Carlo scheme applied, a posteriori, on the swarm of trajectories evaluated without losses. The disappearance of the photon from the cavity is included as a stochastic event whose realization is evaluated at each time step. Each of the 300 original trajectories are replicated five times, and the resulting set of 1500 trajectories is rerun. The probability of photon loss $p_{cav}(t)$ at a given time ($t$) is evaluated as:

$$p_{\mathrm{cav}} = \frac{1}{\tau_c}\Delta t \left| C^{\Gamma}_{S_0,1}(t) \right|^2 = \frac{1}{\tau_c}\Delta t P^{\Gamma}_{S_0,1}, \tag{4}$$

where $\Gamma$ is the current polaritonic state for the current trajectory, $\tau_c$ is the photon lifetime in the cavity, and $\Delta t$ is the timestep of the dynamics. The population $P^{\Gamma}_{S_0,1}$ of the $|S_0, 1\rangle$ uncoupled state in the current polaritonic state $\Gamma$ is defined as the square modulus of $C^{\Gamma}_{S_0,1}$, which is one of the expansion coefficients of the current state $\Gamma$ on the uncoupled basis. In each step, this probability is evaluated and a uniform random number in the interval [0, 1] is generated. If such a number is lower than the decay probability, the total wavefunction for the trajectory is collapsed on the $|S_0, 0\rangle$ state, and the trajectory is stopped. Compared to the direct inclusion of the photon decay in the dynamics, this procedure is neglecting the (very unlikely) event that once in the $|S_0, 1\rangle$ state, the system fluctuates back into the higher energy $|S_1, 0\rangle$ state (or rather the corresponding polaritonic state).

**Code availability**. The calculations were based on a locally modified version of MOPAC2002, which is available from G.G. and M.P. upon reasonable request.

## Data availability

The data that support the findings of this study are available from the open Zenodo repository https://doi.org/10.5281/zenodo.1423796. Additional data are available from the corresponding authors upon reasonable request.

## Acknowledgements

J.F., E.C. and S.C. acknowledge funding from the ERC under the grant ERC-CoG-681285 TAME- Plasmons. G.G. acknowledges funding from the University of Pisa, PRA 2017 28.

## Author contributions

S.C. initiated this project; J.F., G.G., M.P. and S.C. designed the investigation; J.F. and G.G. extended the specific surface hopping scheme to polaritonic states and implemented the software modifications; J.F. performed the calculations; all the authors contributed to the analysis of results and to the writing of the paper.

## Additional information

**Supplementary Information** accompanies this paper at https://doi.org/10.1038/s41467-018-06971-y.

**Competing interests:** The authors declare no competing interests.